\documentclass[a4paper,12pt]{article}

\usepackage{indentfirst}

\usepackage{amsfonts,amsmath,amssymb,bm,a4wide,graphicx,cite,makeidx,multicol}

\fussy

\begin{document}

\title{Quantum Theory of Neutrino Spin-Light in Matter}

\author{A.Grigoriev\footnote{ax.grigoriev@mail.ru}, \
        A.Studenikin \footnote{studenik@srd.sinp.msu.ru },
   \\
   \small {\it Department of Theoretical Physics,}
   \\
   \small {\it Moscow State University,}
   \\
   \small {\it 119992 Moscow,  Russia }
   \\
   A.Ternov \footnote {a\_ternov@mail.ru}
   \\
   \small{\it Department of Theoretical Physics,}
   \\
   \small {\it Moscow Institute for Physics and Technology,}
   \\
   \small {\it 141700 Dolgoprudny, Russia }}

\date{}
\maketitle

\begin{abstract}
    The quantum theory of the spin light of neutrino ($SL\nu$)
    exactly accounting for the effect of the background matter
    is developed. The $SL\nu$ rate and power, and also the emitted
    photon's energy are obtained for the different values of the
    neutrino momentum and density of matter. The spatial
    distribution of the radiation power and the dependence of the
    emitted photon's energy on the direction
    of radiation are studied. It is also shown that,
    in a wide range of the neutrino momentum
    and densities of matter, the $SL\nu$
    radiation is characterized by nearly
    total circular polarization. Conditions for the effective neutrino
    spin-light radiation in the electron plasma are discussed.
\end{abstract}

\section {Introduction}
It is well known that a neutrino with non-zero mass may have
non-trivial electromagnetic properties. In particular, the Dirac
massive neutrino can posses non-vanishing magnetic and electric
dipole moments.  It is believed that non-zero neutrino magnetic
moment could have an important impact on astrophysics and
cosmology.

In the minimally extended Standard Model with $SU(2)$-singlet
right-handed neutrino the one-loop radia\-tive correction
generates neutrino magnetic moment which is proportional to the
neutrino mass \cite{FujShr80}
\begin{equation}
\mu_{\nu}=\frac{3}{8 \sqrt{2}\pi^{2}}eG_{F}m_{\nu}=3 \times 10^{-
19}\mu_{0}\bigg({m_{\nu} \over {1 \mathrm{eV}}}\bigg),
\end{equation}
where $\mu_{0}=e/2m$ is the Bohr magneton, $m_{\nu}$ and $m$ are
the neutrino and electron masses, respectively. There are also
models \cite{KBMVVFY76-87} in which much large values for magnetic
moments of neutrinos are predicted.

So far, the most stringent laboratory constraints on the electron,
muon, and tau neutrino magnetic moments come from elastic
neutrino-electron scattering expe\-ri\-ments. The now days
constraints for the magnetic moments of the three flavour
neutrinos are as follows \cite{RevParPhy02}:  $\mu_{\nu_{e}} \leq
1.5 \times 10^{-10} \mu_{0}, \ \ \mu_{\nu_{\mu}} \leq 6.8 \times
10^{-10} \mu_{0},\ \ \mu_{\nu_{\tau}} \leq 3.9 \times 10^{-10}
\mu_{0}$ .

The recent studies of a massive neutrino elec\-tro\-magnetic
properties within the one-loop level, including the neutri\-no
magnetic moment, can be found in \cite{DvoStuPRD04}, the detailed
discussion on the neutrino charge radius is also presented in
\cite{FujShrPRD04, BerPapBin0405288}. Electromagnetic properties
of a massive neutrino in external elec\-tro\-magnetic fields were
studied in \cite{BorZhukKurTerSJNP85_89EgoLikStu_ICAS99} and in
the background matter were discussed in \cite{NiePal89} (see also
\cite{BhaPal02} for a review).

In a series of papers \cite{LobStuPLB03_04, DvoGriStuIJMPD04} we
have predicted that a massive neutrino moving in matter and
external electromagnetic and/or gravitational fields can emit
electromagnetic radiation due the neutrino non-zero magnetic
moment. We have named this radiation the "spin light of neutrino"
($SL\nu$) in matter and external fields. The quasi-classical
theory of the $SL\nu$ in matter  was developed in
\cite{LobStuPLB03_04, DvoGriStuIJMPD04} on the basis of
Bargmann-Michel-Telegdi equation generalized \cite{EgoLobStuPLB00,
LobStuPLB01, DvoStuJHEP02, StuYF04Stu0407010} for the case of a
neutrino moving in external fields. In our recent papers
\cite{StuTerPLB05_0410296_97} we have developed the quantum
approach to the $SL\nu$ in matter. However, at the final stage of
the performed calculations of the the $SL\nu$ rate and total power
we have considered the limit of low matter density. Therefore, the
evaluation of the consistent quantum theory of the $SL\nu$ in
matter has still remained an open issue. Note that the neutrino
radiation due to the magnetic moment interaction with a constant
magnetic field was considered in \cite{BorZhuTer88}.

Here below we present the quantum theory of the neutrino
spin-light exactly accounting for the effect of the background
matter on the basis of the obtained \cite{StuTerPLB05_0410296_97}
exact solutions of the modified Dirac equation for the neutrino
wave function in matter.

\section{Neutrino wave function in matter}

For developing of the quantum theory of the $SL\nu$ in matter one
has to find the neutrino wave functions with the effect of matter
accounted for. To derive the quantum equation for the neutrino
wave function in the background matter we start with the effective
Lagrangian that describes the neutrino interaction with particles
of the background matter. For definiteness, we consider the case
of the electron neutrino $\nu$ propagating through moving and
polarized matter composed of only electrons (the electron gas).
Assume that the neutrino interactions are described by the
extended standard model supplied with $SU(2)$-singlet right-handed
neutri\-no $\nu_{R}$. We also suppose that there is a macroscopic
amount of electrons in the scale of a neutrino de Broglie wave
length. Therefore, the interaction of a neutrino with the matter
(electrons) is coherent. In this case the averaged over the matter
electrons addition to the vacuum neutrino Lagrangian, accounting
for the charged and neutral interactions, can be written in the
form
\begin{equation}\label{Lag_f}
\begin{array}{c}
  \Delta L_{eff}=-f^\mu \Big(\bar \nu \gamma_\mu {1+\gamma^5 \over
2} \nu \Big),
\\ \\
  f^\mu={G_F \over \sqrt2}\Big((1+4\sin^2 \theta
_W) j^\mu - \lambda ^\mu \Big),
\end{array}
\end{equation}
where the electrons current $j^{\mu}$ and electrons polarization
$\lambda^{\mu}$ are given by
\begin{equation}
j^\mu=(n,n{\bf v}), \label{j}
\end{equation}
and
\begin{equation} \label{lambda}
\lambda^{\mu} =\Bigg(n ({\bm \zeta} {\bf v} ), n {\bm \zeta}
\sqrt{1-v^2}+ {{n {\bf v} ({\bm \zeta} {\bf v} )} \over
{1+\sqrt{1- v^2}}}\Bigg),
\end{equation}
$\theta _{W}$ is the Weinberg angle.

 The Lagrangian (\ref{Lag_f})
accounts for the possible effect of the matter motion and
polarization. Here $n$, ${\bf v}$, and ${\bm \zeta} \ (0\leqslant
|{\bm \zeta} |^2 \leqslant 1)$ denote, respectively, the number
density of the background electrons, the speed of the reference
frame in which the mean momenta of the electrons is zero, and the
mean value of the polarization vector of the background electrons
in the above men\-tioned reference frame. The detailed discussion
on the determina\-tion of the electrons polarization can be found
in \cite{LobStuPLB01, DvoStuJHEP02, StuYF04Stu0407010}.

From the standard model Lagrangian with the extra term $\Delta
L_{eff}$ being added, we derive the following modi\-fied Dirac
equation for the neutrino moving in the background matter,
\begin{equation}\label{new} \Big\{
i\gamma_{\mu}\partial^{\mu}-\frac{1}{2}
\gamma_{\mu}(1+\gamma_{5})f^{\mu}-m \Big\}\Psi(x)=0.
\end{equation}
This is the most general equation of motion for a neutrino in
which the effective potential
$V_{\mu}=\frac{1}{2}(1+\gamma_{5})f_{\mu}$ accounts for both the
charged and neutral-current interactions with the background
matter and also for the possible effects of the matter motion and
polarization.

If we neglect the contribution of the neutral-current interaction
and possible effects of motion and polariza\-tion of the matter
then from (\ref{new}) we can get corres\-ponding equations for the
left-handed and right-handed chiral components of the neutrino
field derived in \cite{PanPLB91-PRD92}.

The generalizations of the modified Dirac equation (\ref{new}) for
more complicated matter compositions (and the other flavour
neutrinos) are just straightforward. For instance, if one
considers the case of realistic matter composed of electrons,
protons and neutrons, then the matter term in (\ref{new}) is (see,
for instance, \cite {PalIJMPA92}, \cite{DvoStuJHEP02} and the
second paper of \cite{StuTerPLB05_0410296_97})
\begin{equation}\label{f_mu}
f^\mu={G_F \over \sqrt2}\sum\limits_{f=e,p,n}
j^{\mu}_{f}q^{(1)}_{f}+\lambda^{\mu}_{f}q^{(2)}_{f}
\end{equation}
where
\begin{align*}\label{q_f}
  q^{(1)}_{f}= &
  (I_{3L}^{(f)}-2Q^{(f)}\sin^{2}\theta_{W}+\delta_{ef}), \\
  q^{(2)}_{f}= &
  -(I_{3L}^{(f)}+\delta_{ef}),
\end{align*}
$$
\delta_{ef}=
  \begin{cases}
    1 & \text{for} {\it f=e}, \\
    0 & \text{for} {\it f=n, p}.
  \end{cases}
$$
and $I_{3L}^{(f)}$ is the value of the isospin third component of
a fermion $f$, $Q^{(f)}$ is the value of its electric charge and
$\theta_{W}$ is the Weinberg angle. The fermion's currents
$j^{\mu}_{f}$ and polarizations $\lambda^{\mu}_{f}$ are given by
eqs.(\ref{j}) and (\ref{lambda}) with the appropriate
substitutions: $n, \bm \zeta, {\bf v} \rightarrow n_f, {\bm
\zeta}_f, {\bf v}_f$.

In the further discussion below we consider the case when no
electromagnetic field is present in the background. We also
suppose that the matter is un\-polarized, $\lambda^{\mu}=0$.
Therefore, the term describing the neutrino interaction with the
matter is given by
\begin{equation}\label{f}
f^\mu=\frac{\tilde{G}_{F}}{\sqrt2}(n,n{\bf v}),
\end{equation}
where we use the notation $\tilde{G}_{F}={G}_{F}(1+4\sin^2 \theta
_W)$.

In the rest frame of the matter the equation (\ref{new}) can be
written in the Hamiltonian form,
\begin{equation}\label{H_matter}
i\frac{\partial}{\partial t}\Psi({\bf r},t)=\hat H_{matt}\Psi({\bf
r},t),
\end{equation}
where
\begin{equation}\label{H_G}
  \hat H_{matt}=\hat {\bm{\alpha}} {\bf p} + \hat {\beta}m +
  \hat V_{matt},
\end{equation}
and
\begin{equation}\label{V_matt}
\hat V_{matt}= \frac{1}{2\sqrt{2}}(1+\gamma_{5}){\tilde {G}}_{F}n,
\end{equation}
here $\bf p$ is the neutrino momentum. We use the Pauli-Dirac
representation for the Dirac matrices $\hat {\bm \alpha}$ and
$\hat {\beta}$, in which
\begin{equation}\label{a_b}
   {\hat {\bm \alpha}}=
\left(
   \begin{array}{cc}
   0 & \hat {\bm {\sigma}} \\
   {\hat {\bm {\sigma}}} & 0
   \end{array}
\right) =\gamma_0{\bm \gamma}, \ \ \ {\hat \beta}= \left(
   \begin{array}{cc}
   1 & 0 \\
   0 & -1
   \end{array}
\right) =\gamma_0,
\end{equation}
where ${\hat { \bm\sigma}}=({ \sigma}_{1},{ \sigma}_{2},{
\sigma}_{3})$ are the Pauli matrixes.

The form of the Hamiltonian (\ref{H_G}) implies that the operators
of the momentum, $\hat {\bf p}$, and longitudinal
po\-lariza\-tion, ${\hat{\bf \Sigma}} {\bf p}/p$, are the
integrals of motion. So that, in particular, we have
\begin{equation}\label{helicity}
  \frac{{\hat{\bf \Sigma}}{\bf p}}{p}
  \Psi({\bf r},t)=s\Psi({\bf r},t),
 \ \ {\hat {\bf \Sigma}}=
\left(
   \begin{array}{cc}
   {\hat {\bm \sigma}} & 0 \\
   0 & {\hat {\bm \sigma}}
   \end{array}
\right),
\end{equation}
where the values $s=\pm 1$ specify the two neutrino helicity
states, $\nu_{+}$ and  $\nu_{-}$. In the relativistic limit the
negative-helicity neutrino state is dominated by the left-handed
chiral state ($\nu_{-}\approx \nu_{L}$), whereas the
positive-helicity state is dominated by the right-handed chiral
state ($\nu_{+}\approx \nu_{R}$).

For the stationary states of the equation (\ref{new}) we get
\begin{equation}\label{stat_states}
\Psi({\bf r},t)=e^{-i(  E_{\varepsilon}t-{\bf p}{\bf r})}u({\bf
p},E_{\varepsilon}),
\end{equation}
where $u({\bf p},E_{\varepsilon})$ is independent on the
coordinates and time. Upon the condition that the equation
(\ref{new}) has a non-trivial solution, we arrive to the energy
spectrum of a neutrino moving in the background matter:
\begin{equation}\label{Energy}
  E_{\varepsilon}=\varepsilon{\sqrt{{\bf p}^{2}\Big(1-s\alpha \frac{m}{p}\Big)^{2}
  +m^2} +\alpha m} ,
\end{equation}
where we use the notation
\begin{equation}\label{alpha}
  \alpha=\frac{1}{2\sqrt{2}}{\tilde G}_{F}\frac{n}{m}.
\end{equation}
The quantity $\varepsilon=\pm 1$ splits the solutions into the two
branches that in the limit of the vanishing matter density,
$\alpha\rightarrow 0$, reproduce the positive and
negative-frequency solutions, respectively. It is also important
to note that the neutrino energy in the background matter depends
on the state of the neutrino longitudinal polarization, i.e. in
the relativistic case the left-handed and right-handed neutrinos
with equal momenta have different energies.

The procedure, similar to one used for derivation of the solution
of the Dirac equation in vacuum, can be adopted for the case of a
neutrino moving in matter. We apply this procedure to the equation
(\ref{new}) and arrive to the final form of the wave function of a
neutrino moving in the background matter:
\begin{equation}\label{wave_function}
  \Psi_{\varepsilon,{\bf p},s}({\bf r},t) =
\frac{e^{-i( E_{\varepsilon}t-{\bf p}{\bf r})}}{2L^{\frac{3}{2}}}
     \left(
   \begin{array}{cccc}
      {\sqrt{1+ \frac{m}{E_{\varepsilon}-\alpha m}}}
      \sqrt{1+s\frac{p_3}{p}}
      \\
      {s \sqrt{1+ \frac{m}{ E_{\varepsilon}-\alpha m}}}
      \sqrt{1-s\frac{p_3}{p}}\ \ e^{i\delta}
      \\
      {s\varepsilon\sqrt{1- \frac{m}{ E_{\varepsilon}-\alpha m}}}
      \sqrt{1+s\frac{p_3}{p}}
      \\
      {\varepsilon\sqrt{1- \frac{m}{ E_{\varepsilon}-\alpha m}}}
      \sqrt{1-s\frac{p_3}{p}}\ e^{i\delta}
   \end{array}
\right)
\end{equation}
where the energy $E_{\varepsilon}$ is given by (\ref{Energy}),$L$
is the nor\-malization length and $\delta=\arctan{p_2/p_1}$. In
the limit of vanishing density of matter, when $\alpha\rightarrow
0$, the wave function (\ref{wave_function}) transforms to the
vacuum solution of the Dirac equation.

The quantum equation (\ref{new}) for a neutrino in the background
matter with the obtained exact solution (\ref{wave_function}) and
energy spectrum (\ref{Energy}) establish a basis for a very
effective method (similar to the Furry repre\-senta\-tion of
quantum electrodynamics) in investigations of different phenomena
that can appear when neutrinos are moving in the media.

Let us now consider in some detail the properties of a neutrino
energy spectrum (\ref{Energy}) in the background matter that are
very important for understanding of the mechanism of the neutrino
spin light phenomena. For the fixed magnitude of the neutrino
momentum $p$ there are the two values for the "positive sign"
($\varepsilon =+1$) energies
\begin{equation}\label{Energy_nu}
\begin{array}{c}
   E^{s=+1}={\sqrt{{\bf p}^{2}\Big(1-\alpha \frac{m}{p}\Big)^{2}
   +m^2} +\alpha m}, \\
   E^{s=-1}={\sqrt{{\bf p}^{2}\Big(1+\alpha \frac{m}{p}\Big)^{2}
   +m^2} +\alpha m},
\end{array}
\end{equation}
that determine the positive- and negative-helicity eigen\-states,
respectively. The energies in Eq.(\ref{Energy_nu}) correspond to
the particle (neutrino) solutions in the background matter. The
two other values for the energy, correspon\-ding to the negative
sign $\varepsilon =-1$, are for the antiparticle solutions. As
usual, by changing the sign of the energy, we obtain the values
\begin{equation}\label{Energy_anti_nu}
\begin{array}{c}
  {\tilde E}^{s=+1}={\sqrt{{\bf p}^{2}
  \Big(1-\alpha \frac{m}{p}\Big)^{2}
  +m^2} -\alpha m}, \\
  {\tilde E}^{s=-1}={\sqrt{{\bf p}^{2}
  \Big(1+\alpha \frac{m}{p}\Big)^{2}
  +m^2} -\alpha m},
\end{array}
\end{equation}
that correspond to the positive- and negative-helicity
antineutrino states in the matter. The expressions in
Eqs.(\ref{Energy_nu}) and (\ref{Energy_anti_nu}) would reproduce
the neutrino disper\-sion relations of \cite{PanPLB91-PRD92} (see
also \cite{WeiKiePRD97}), if the contribution of the
neutral-current interaction to the neutrino potential were left
out.

\section {Quantum theory of $SL\nu$ with exact account for
matter density} In this section we should like to use the obtained
solu\-tions (\ref{wave_function}) of the equation (\ref{new}) for
a neutrino moving in the background matter for the study of the
spin light of neutrino ($SL\nu$) in the matter. We develop below
the {\it quantum} theory of this effect exactly accounting for the
matter number density $n$. Note that in our previous studies
\cite{StuTerPLB05_0410296_97} the $SL\nu$ rate and radiation power
were derived in the limit of not very high densities of the
background matter (see also our comments to Eq.(\ref{gamma_1})
below).

Within the quantum approach, the corresponding Feynman diagram of
the $SL\nu$ in the matter is the standard one-photon emission
diagram with the initial and final neutrino states described by
the "broad lines" \ that account for the neutrino interaction with
the matter. It follows from the usual neutrino magnetic moment
interaction with the quantized photon field, that the amplitude of
the transition from the neutrino initial state $\psi_{i}$ to the
final state $\psi_{f}$, accompanied by the emission of a photon
with a momentum $k^{\mu}=(\omega,{\bf k})$ and  polarization ${\bf
e}^{*}$, can be written in the form
\begin{equation}\label{amplitude}
  S_{f i}=-\mu \sqrt{4\pi}\int d^{4} x {\bar \psi}_{f}(x)
  ({\hat {\bf \Gamma}}{\bf e}^{*})\frac{e^{ikx}}{\sqrt{2\omega L^{3}}}
   \psi_{i}(x),
\end{equation}
where $\mu$ is the neutrino magnetic moment, $\psi_{i}$ and
$\psi_{f}$ are the corresponding exact solutions of the equation
(\ref{new}) given by (\ref{wave_function}), and
\begin{equation}\label{Gamma}
  \hat {\bf \Gamma}=i\omega\big\{\big[{\bf \Sigma} \times
  {\bm \varkappa}\big]+i\gamma^{5}{\bf \Sigma}\big\}.
\end{equation}
Here ${\bm \varkappa}={\bf k}/{\omega}$ is the unit vector
pointing the direction of the emitted photon propagation. Note
that the deve\-loped approach to the considered process in the
presence of matter is similar to the Furry representation in
studies of quantum processes in external electromagnetic fields.

The integration in (\ref{amplitude}) with respect to time yields
\begin{equation}\label{amplitude}
   S_{f i}=
  -\mu {\sqrt {\frac {2\pi}{\omega L^{3}}}}
  ~2\pi\delta(E_{f}-E_{i}+\omega)
  \int d^{3} x {\bar \psi}_{f}({\bf r})({\hat {\bf \Gamma}}{\bf e}^{*})
  e^{i{\bf k}{\bf r}}
   \psi_{i}({\bf r}),
\end{equation}
where the delta-function stands for the energy con\-ser\-vation.
Performing the integrations over the spatial co-ordinates, we can
recover the delta-functions for the three com\-ponents of the
momentum. Finally, we get the law of the energy-momentum
conservation for the considered process,
\begin{equation}\label{e_m_con}
    E_{i}=E_{f}+\omega, \ \ \
    {\bf p}_{i}={\bf p}_{f}+{\bm \varkappa}.
\end{equation}
where $E_{i}$ and $E_{f}$ are the energies of the initial and
final neutrino states in matter. From (\ref{e_m_con}) it follows
that the emitted photon energy $\omega$ exhibits the critical
dependence on the helicities of the initial and final neutrino
states. In the case of electron neutrino moving in  matter
composed of electrons $\alpha$ is positive. Thus, it follows that
the only possibility for the $SL\nu$ to appear is provided in the
case when the neutrino initial and final states are characterized
by $s_{i}=-1$ and $s_{f}=+1$, respectively. We conclude that in
the considered process the relativistic left-handed neutrino is
converted to the right-handed neutrino. A discussion on the main
properties of the $SL\nu$ emitted by different flavour neutrinos
moving in matter composed of electrons, protons and neutrons can
be found in \cite{StuTerPLB05_0410296_97} ( see also
 \cite{Lob_hep_ph_0411342}).

The emitted photon energy in the considered case ($s_i=-s_f=-1$)
obtained as an exact solution of the equations (\ref{e_m_con})
reads
\begin{equation}\label{omega1}
\omega =\frac{2\alpha mp\left[ (E-\alpha m)-\left( p+\alpha
m\right) \cos \theta \right] }{\left( E-\alpha m-p\cos \theta
\right) ^{2}-\left( \alpha m\right) ^{2}},
\end{equation}
where $\theta$ is the angle between ${\bm \varkappa}$ and the
direction of the initial neutrino propagation.

In the case of not very high density of matter, when the parameter
${\tilde {G}}_{F}n/m\ll 1$,  we can expand the photon energy
(\ref{omega1}) over the above mentioned parameter and in the liner
approximation get the result of \cite{StuTerPLB05_0410296_97}:
\begin{equation}\label{omega_2}
    \omega=
    \frac {1}{1-\beta \cos
    \theta}\omega_0,
\end{equation}
where
\begin{equation}\label{omega_0}
\omega_0= \frac {{\tilde G}_{F}} {\sqrt{2}}n\beta,
\end{equation}
and $\beta$ is the neutrino speed in vacuum.

Using the wave functions (\ref{wave_function}) for the neutrino
initial and final states in matter we calculate the spin light
transition rate exactly accounting for the matter density
parameter and get
\begin{equation}\label{Gamma}
 \Gamma =\mu ^{2}\int_{0}^{\pi
}\omega ^{3} \big[(\tilde\beta \tilde\beta ^{\prime }+1)(1-y\cos
\theta )-(\tilde\beta +\tilde\beta ^{\prime }) (\cos \theta
-y)\big]\frac{\sin \theta }{1+\tilde\beta ^{\prime }y} d\theta ,
\end{equation}
Here we use the notations
\begin{equation}
\tilde \beta =\frac{p+\alpha m}{E-\alpha m}, \ \ \tilde \beta
^{\prime }=\frac{p^{\prime }-\alpha m}{E^{\prime }-\alpha m},
\end{equation}
where the final neutrino energy and momentum are
\begin{equation}
E^{\prime }=E-\omega , \ \ \ p^{\prime }=K\omega -p,
\end{equation}
respectively, and
\begin{equation}
y=\frac{\omega -p\cos \theta }{p^{\prime }}, \ \ K=\frac{E-\alpha
m-p\cos \theta }{\alpha m}.
\end{equation}

Performing the integration in Eq.(\ref{Gamma}), we obtain for the
$SL\nu$ rate in matter
\begin{eqnarray}\label{gamma}
\Gamma &=&\frac{1}{2\left( E-p\right) ^{2}\left( E+p-2\alpha
m\right) ^{2}\left( E-\alpha m\right) p^{2}} \notag
\\
&&\times \left\{ \left( E^{2}-p^{2}\right) ^{2}\left(
p^{2}-6\alpha ^{2}m^{2}+6E\alpha m-3E^{2}\right) \left( \left(
E-2\alpha m\right) ^{2}-p^{2}\right) ^{2}\right. \notag
\\
&&\times \ln \left[ \frac{\left( E+p\right) \left( E-p-2\alpha
m\right) }{\left( E-p\right) \left( E+p-2\alpha m\right) }\right]
+4\alpha mp\left[ 16\alpha ^{5}m^{5}E\left( 3E^{2}-5p^{2}\right)
\right. \notag
\\
&&-8\alpha ^{4}m^{4}\left( 15E^{4}-24E^{2}p^{2}+p^{4}\right)
+4\alpha ^{3}m^{3}E\left( 33E^{4}-58E^{2}p^{2}+17p^{4}\right)
\notag
\\
&&-2\alpha ^{2}m^{2}\left( 39E^{2}-p^{2}\right) \left(
E^{2}-p^{2}\right) ^{2}+12\alpha mE\left( 2E^{2}-p^{2}\right)
\left( E^{2}-p^{2}\right) ^{2} \notag
\\
&&-\left. \left. \left( 3E^{2}-p^{2}\right) \left(
E^{2}-p^{2}\right) ^{3} \right] \right\},
\end{eqnarray}
where the energy of the initial neutrino is given by
(\ref{Energy}).

As it follows from (\ref{gamma}), the $SL\nu$ rate is a rather
complicated function of the neutrino momentum $p$ and mass $m$, it
is also non-trivially dependent on the matter density parameter
$\alpha$. In the limit of low densities of matter, $\alpha\ll 1$,
we get
\begin{equation}\label{gamma_1}
\Gamma \simeq \frac{64}{3}\frac{\mu^2 \alpha ^{3}p^{3}m}{E_{0}},
\end{equation}
where $E_{0}=\sqrt{p^{2}+m^{2}}$. The obtained expression is in
agreement with our results of refs.\cite{LobStuPLB03_04,
StuTerPLB05_0410296_97}. Note that the considered limit of
$\alpha\ll 1$ can be appropriate even for very dense media of
neutron stars with $n\sim 10^{33} \ \ cm^{-3}$ because
$\frac{1}{2\sqrt{2}}{\tilde G}_{F}n\sim 1 \ eV$ for a medium
characterized by $n=10^{37} \ \ cm^{-3}$.

Let us consider the $SL\nu$ rate for the different limiting values
of the neutrino momentum $p$ and matter density parameter
$\alpha$. In the relativistic case $p\gg m$ from (\ref{gamma}) we
get
\begin{equation}
\Gamma = \left\{
  \begin{tabular}{cc}
  \ $\frac{64}{3} \mu ^2 \alpha ^3 p^2 m,$ &
  \ \ \ for {$\alpha \ll \frac{m}{p},$ } \\
  \ $4 \mu ^2 \alpha ^2 m^2 p$, & \ \ \ \ \ \ \ \
  { for
  $ \frac{m}{p} \ll \alpha \ll \frac{p}{m},$} \\
\ $4 \mu ^2 \alpha ^3 m^3$, & \ { for
  $ \alpha \gg \frac{p}{m}. $}
  \end{tabular}
\right.
\end{equation}
In the opposite case $p\ll m$ we get
\begin{equation}
\Gamma = \left\{
  \begin{tabular}{cc}
  \ $\frac{64}{3} \mu ^2 \alpha ^3 p^3,$ &
  \ \ \ for {$\alpha \ll 1,$ } \\
  \ $\frac{512}{5} \mu ^2 \alpha ^6 p^3$, & \ \ \ \ \ \ \ \  { for
  $ 1 \ll \alpha \ll \frac{m}{p},$} \\
\ $4 \mu ^2 \alpha ^3 m^3$, & \ { for
  $ \alpha \gg \frac{m}{p}. $}
  \end{tabular}
\right.
\end{equation}

On the basis of (\ref{amplitude}) we also derive the $SL\nu$
radiation power:
\begin{equation}\label{power}
I=\mu ^2\int_{0}^{\pi }\omega ^{4}\big[(\tilde\beta \tilde\beta
^{\prime }+1)(1-y\cos \theta )-(\tilde\beta +\tilde\beta ^{\prime
})(\cos \theta -y)\big]\frac{\sin \theta }{1+\tilde\beta
^{\prime}y}d\theta.\notag
\end{equation}
Performing the integration and considering the case $\alpha\ll 1$,
we get for the total power
\begin{equation}\label{power_1}
I\simeq \frac{128}{3}\mu ^{2}\alpha ^{4}p^{4}
\end{equation}
in agreement with refs.\cite{LobStuPLB03_04,
StuTerPLB05_0410296_97}.

Let us now consider the $SL\nu$ radiation power for the different
limiting values of the neutrino momentum $p$ and matter density
parameter $\alpha$. In the relativistic case $p\gg m$ from
(\ref{power}) we get
\begin{equation}
I= \left\{
  \begin{tabular}{cc}
  \ $\frac{128}{3}\mu ^{2}\alpha ^{4}p^{4},$ &
  \ \ \ for {$\alpha \ll \frac{m}{p},$ } \\
  \ $\frac{4}{3} \mu ^2 \alpha ^2 m^2 p^2$, & \ \ \ \ \ \ \ \  { for
  $ \frac{m}{p} \ll \alpha \ll \frac{p}{m},$} \\
\ $4 \mu ^2 \alpha ^4 m^4$, & \ { for
  $ \alpha \gg \frac{p}{m}. $}
  \end{tabular}
\right.
\end{equation}
In the opposite case $p\ll m$ we get
\begin{equation}
I = \left\{
  \begin{tabular}{cc}
  \ $\frac{128}{3} \mu ^2 \alpha ^4 p^4,$ &
  \ \ \ for {$\alpha \ll 1,$ } \\
  \ $\frac{1024}{3} \mu ^2 \alpha ^8 p^4$, & \ \ \ \ \ \ \ \  { for
  $ 1 \ll \alpha \ll \frac{m}{p},$} \\
\ $4 \mu ^2 \alpha ^4 m^4$, & \ { for
  $ \alpha \gg \frac{m}{p}. $}
  \end{tabular}
\right.
\end{equation}

Note that the obtained $SL\nu$ rate and radiation power in the
case $p\gg m$ for $\alpha \gg \frac{m}{p}$  are in agreement with
\cite{Lob_hep_ph_0411342}.

From the obtained expressions for the $SL\nu$ rate and total power
it is possible to get an estimation for the emitted photons
average energy:
\begin{equation}\label{average_energy}
\left\langle \omega\right\rangle = \frac{I}{\Gamma}.
\end{equation}

 In the case $p\gg m$ we get
\begin{equation}
\left\langle \omega\right\rangle \simeq \left\{
  \begin{tabular}{cc}
  \ $2\alpha \frac{p^{2}}{m},$ &
  \ \ \ for {$\alpha \ll \frac{m}{p},$ } \\
  \ $\frac{1}{3} p $, & \ \ \ \ \ \ \ \  { for
  $ \frac{m}{p} \ll \alpha \ll \frac{p}{m},$} \\
\ $\alpha m$, & \ { for
  $ \alpha \gg \frac{p}{m}. $}
  \end{tabular}
\right.
\end{equation}
In the opposite case $p\ll m$ for the emitted photons overage
energy  we have
\begin{equation}
\left\langle \omega\right\rangle \simeq\left\{
  \begin{tabular}{cc}
  \ $2 p\alpha ,$ &
  \ \ \ for {$\alpha \ll 1,$ } \\
  \ $\frac{10}{3}p\alpha^2$, & \ \ \ \ \ \ \ \
{ for
  $ 1 \ll \alpha \ll \frac{m}{p},$} \\
\ $\alpha  m $, & \ { for
  $ \alpha \gg \frac{m}{p}. $}
  \end{tabular}
\right.
\end{equation}

\section {Conclusion}
We should like to note that for a wide range of the neutrino
momentum $p$ and density parameter $\alpha$ the $SL\nu$ power is
collimated along the direction of the neutrino propagation. The
form of the radiation power spatial distribution calculated with
use of Eq.(\ref{power}) in the case $p > m$ for low and high
density of matter  are shown in Figs. 1 and 2, respectively. As it
follows from these figures, the form of the distribution depends
on the density of matter. The form of the spatial distribution of
the radiation moves from the projector-like shape to the cap-like
one with increase of the matter density. From (\ref{power}) we
derive, that in the case of $p\gg m$ for a wide range of the
matter densities, $\alpha\ll \frac{p}{m}$, the direction of the
maximum in the spatial distribution of the radiation power is
characterized by the angle
\begin{equation}\label{angle_max}
\cos \theta_{max} \simeq 1-\frac{2}{3}\alpha\frac{ m}{p}.
\end{equation}
It follows that in a dense matter the $SL\nu$ radiation in the
direction of the initial neutrino motion is strongly suppressed,
whereas there is a lighted ring on the plane per\-pen\-dicular to
the neutrino motion.

The quantum calculation of the radiation power with the photon's
circular polarization being accounted for (see also
\cite{StuTerPLB05_0410296_97}) shows another very important
property of the $SL\nu$ in matter. In the limit of low matter
density,  $\alpha \ll 1$, we get for the power
\begin{equation}
I^{\left( l\right) }\simeq \frac{64}{3}\mu^{2}\alpha
^{4}p^{4}\left( 1-l\frac{p}{2E_{0}}\right),
\end{equation}
where $l=\pm 1$ corresponds to the right and left photon circular
polarizations, respectively. In this limiting case the power of
radiation of the left-polarized photons exceeds that of the
right-polarized photons. In particular, this result is also valid
for the non-relativistic case $p\ll m$ for low density with
$\alpha \ll 1$. It is remarkable that in the most interesting case
of a rather dense matter ($\alpha \gg \frac{m}{p}$ for $p\gg m $
and $\alpha \gg 1$ for $p\ll m$), the main contribution to the
power is provided by the right-polarized photons, whereas the
emission of the left-polarized photons is suppressed:
\begin{eqnarray}
I^{\left( +1\right) } &\simeq &I, \\
I^{\left( -1\right) } &\simeq &0.
\end{eqnarray}
We conclude that in a dense matter the $SL\nu$ photons are emitted
with nearly total right-circular polarization.

Finally, we should like to discuss in some detail restrictions on
the propagation of the $SL\nu$ photons that are set by the
presence of the background electron plasma in the case of $p\gg m$
for the density parameter  $\frac{m}{p} \ll \alpha \ll
\frac{p}{m}$. Only the photons with energy that exceeds the
plasmon frequency
\begin{equation}\label{pl_freq} \omega _{pl}= \sqrt{\frac{4 \pi
e^2}{m_e} n},
\end{equation}
can propagate in the plasma (here $e^2=\alpha_{QED}$ is the
fine-structure constant and $m_e$ is the mass of the electron).
From (\ref{omega1}) and (\ref{angle_max}) it follows that, on the
one hand, the photon's energy and the radiation power depend on
the direction of the radiation. On the other hand,  we can
conclude that the maximal value of the photon's energy,
\begin{equation}\label{omega_max}
\omega_{max}=p,
\end{equation}
and the  energy the photon emitted in the direction of the maximum
radiation power,
\begin{equation}
\omega (\theta_{max})=\frac{3}{4}p,
\end{equation}
are of the same order in the considered case. For the relativistic
neutrino and rather dense matter  the angle $\theta_{max}$, at
which the radiation power  has its maximum, is very close to zero
when the photon energy reaches the maximal value
(\ref{omega_max}). Therefore, the condition for the effective
$SL\nu$ production and propagation through the plasma can be
written in the form
\begin{equation}\label{p_in_pl}
p\gg p_{min}=3.5\times 10^{4}\left(
\frac{n}{10^{30}cm^{-3}}\right)^{1/2} eV.
\end{equation}
For $n \sim 10^{33} \ cm^{-3}$ we get $p_{min} \sim 1 \ MeV$.

The investigated properties of the neutrino spin light in matter
(i.e., the spatial distribution of the radiation power and the
angular dependence of the emitted photon's energy, as well as
nearly total circular polarization of the radiation) might be
important for the experimental identification of the $SL\nu$ from
different astrophysical and cosmology objects and media.

\begin{figure}[t]
\begin{minipage}{7.5cm}
\includegraphics[scale=.6]{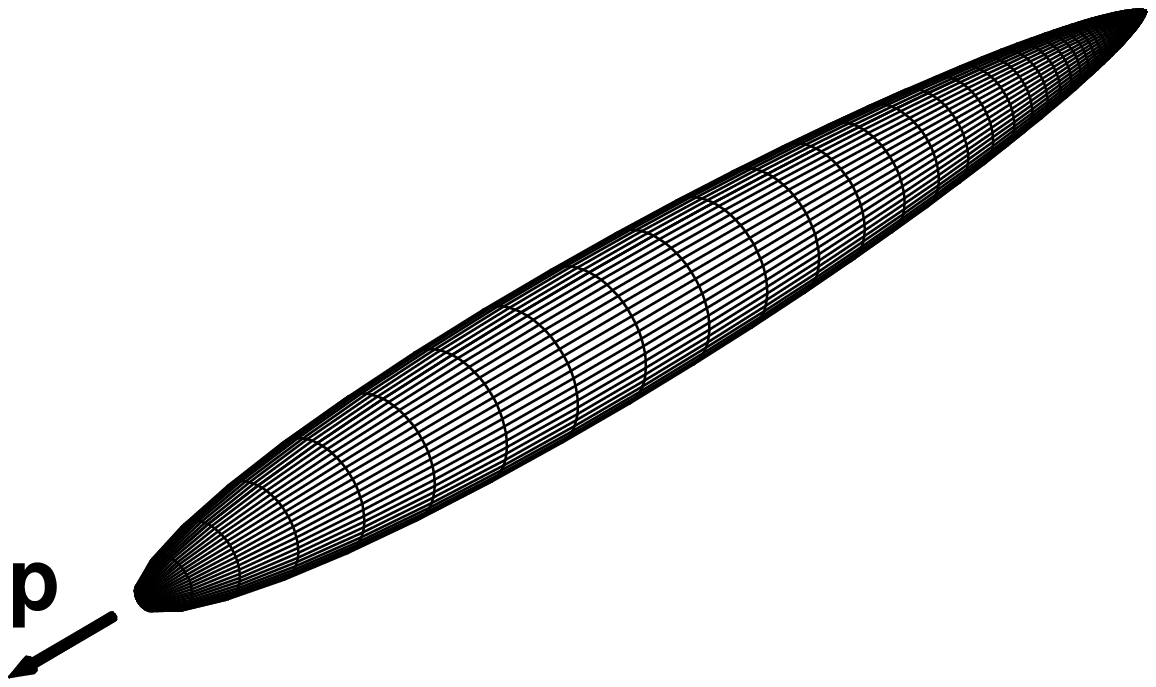}
\caption{The spatial distribution of the $SL\nu$ radiation power
for the case $p/m=5, \ \alpha m=0.01$.}
\end{minipage}
\hfill
\begin{minipage}{7.5cm}
\includegraphics[scale=.6]{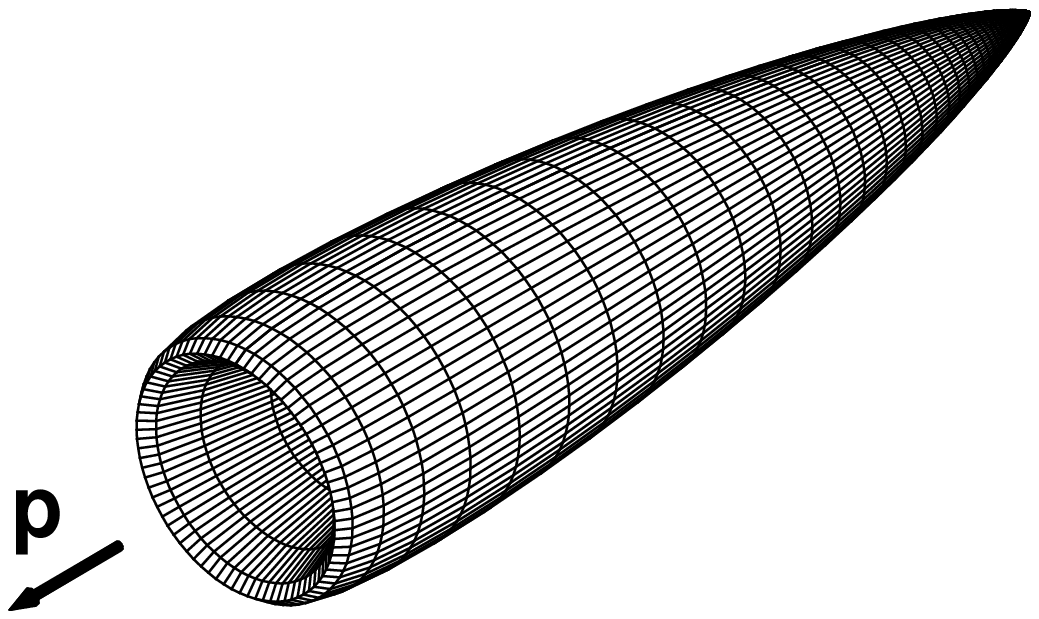}
\caption{The spatial distribution of the $SL\nu$ radiation power
for the case $p/m=10^3, \ \alpha m=100$. }
\end{minipage}
\end{figure}

\end{document}